\documentclass{article}

\usepackage{amssymb}
\usepackage{graphicx}
\usepackage{dcolumn}
\usepackage{bm}
\usepackage{amsmath}
\usepackage{amstext}
\usepackage{amssymb}

\begin{document}

\title{Quantum Hamilton-Jacobi cosmology and classical-quantum correlation}

\author{ M. Fathi$^{1}$\thanks{email: m-fathi@sbu.ac.ir },\hspace{.2cm}and \hspace{.2cm} S. Jalalzadeh$^{1,2}$\thanks{email:
s-jalalzadeh@sbu.ac.ir; shahram.jalalzadeh@unila.edu.br}
\\ $^1${\small Department of Physics, Shahid Beheshti University, G. C. Evin, Tehran 19839, Iran}
\\$^2${\small Federal University of Latin-America Integration, Itaipu Technological Park, PO box 2123,}\\ {\small Foz do Igua\c{}c-PR, 85867-670, Brazil.} }

\maketitle
\begin{abstract}
How the time evolution which is typical for classical cosmology emerges from
quantum cosmology? The answer  is not trivial because the Wheeler-DeWitt equation
is time independent. A framework associating the quantum Hamilton-Jacobi to the minisuperspace cosmological models has been introduced in \cite{New}.  In this paper we show that  time dependence and quantum-classical correspondence both arise naturally in the quantum
Hamilton-Jacobi formalism of quantum mechanics, applied to quantum cosmology. We study the quantum Hamilton-Jacobi cosmology of spatially flat  homogeneous and  isotropic  early universe whose matter content is a perfect
fluid. The classical cosmology emerge around one Planck time
where its linear size is around a few millimeter, without needing any classical inflationary phase afterwards
to make it grow to its present size.
\end{abstract}

\section{Introduction}
In canonical quantum cosmology, the wave function of universe is obtained
from the Wheeler-DeWitt (WDW) equation which is time independent and consequently we have no quantum dynamics.
In this direction, the Copenhagen interpretation
of quantum cosmology  has some serious conceptual  problems: The impossibility of a clear division of the whole universe
into the observer and the observed makes difficult to have a perspicuous physical interpretation for the wave function. Consequently there cannot
be any ``wave function collapse'' as in standard quantum mechanics. Moreover, assuming the existence of only one observable universe, the interpretation
of the absolute square of the wave function as a probability density is impossible.
 To find a solution to the above mentioned problems  the straight and direct way could be the  de Broglie-Bohm (dBB) interpretation of quantum cosmology.
 The dBB interpretation is favorable, especially for a quantum theory of
cosmology, because this interpretation is able to resolve the  conceptual problems of quantum cosmology \cite{Bohmian}.
However, we have a problem  in using  of this interpretation in quantum
cosmology: It cannot describe the  trajectories and non-zero velocities for real wave functions of quantum cosmology.

 In recent years, the Quantum Hamilton-Jacobi (QHJ) formulation of quantum
mechanics has been developed
as a new alternative interpretation of quantum mechanics \cite{Complex} which was developed by Leacock and Padgett  \cite{Leacock}   and independently by Gozzi \cite{Gozzi}.  One of the advantages of this formalism, which follows classical mechanics closely, is that it does not face the problem of stationarity (zero
velocity) of particles in bound states, encountered in the  dBB representation
\cite{John}.
The QHJ formulation can be introduced as follows. We employ $\Psi=e^{iS(q^\mu)}$,
where
$S\,\,\text{and}\,\, q^\mu\in\mathbb{C}$, in the corresponding wave equation of the quantum system to obtain a single QHJ equation \cite{Complex}. Since the  action function $S$ is complex valued the position and conjugate momentum of particles are
also complex valued.
This formalism has been used  to obtain the energy eigenvalues for many
one-dimensional bound state problems and separable problems in higher dimensions
for solvable potentials \cite{R}. Besides the determination of energy eigenvalues, this formalism has been used to obtain quantum trajectories
in complex space.
On the contrary to dBB formalism, the complex quantum trajectory of a particle can
be obtained in the quantum Hamilton-Jacobi formalism for both stationary
and non-stationary states through the QHJ equation and the quantum momentum function,
which is analytically extended to complex space. John \cite{John} has applied this
formalism to several simple analytical examples for time dependent and time
independent
problems. In addition, for stationary states, complex quantum
trajectories satisfying the complex-valued QHJ equation have been analytically studied
for the free particle, the potential step, the potential barrier, the harmonic
potential and the hydrogen atom \cite{1724}.
In this description \cite{Complex}, the transition from a quantum regime
to the corresponding classical world occurs for large values of position and the quantum numbers of system \cite{Har}, where the quantum force  disappears and the particle’s motion is entirely governed by the classical equation of motion.

Recently, in \cite{New} the QHJ formulation of quantum cosmology is developed.
 The authors obtained
the state dependent quantum cosmological solutions with complex trajectories
in the complex minisuperspace. Then it was shown that for large
values of the scale factor and quantum state number $n$ the model emerges into a classical
cosmology without the horizon and flatness problems.

In this paper we will investigate the quantum cosmology of a simple
flat FLRW universe filled with   a perfect fluid in the QHJ framework. In section II, we describe
briefly  the QHJ interpretation of quantum cosmology of minisuperspace models. We obtain the cosmological solutions with complex trajectories in complex minisuperspace and  the quantum-classical correspondence in section III.
Finally, in Section IV, we make our conclusions.
\section{Quantum Hamilton-Jacobi interpretation of minisuperspace  quantum cosmology}
In this section we will review briefly the minisuperspace quantum cosmological
models in the framework
of QHJ
 developed by Fathi, Jalalzadeh and Moniz in  \cite{New}.
Let us start with the line element of  spacetime which is given by
\begin{eqnarray}\label{a1}
ds^2=-N^2(t)dt^2+h_{ij}dx^idx^j,\,\,\,i,j=1,2,3,
\end{eqnarray}
where the 3-metric $h_{ij}$ is restricted to be homogeneous. Then the assumption
of homogeneity of matter fields  leads us to the following Lagrangian
of Einstein-Hilbert plus the matter fields
\begin{eqnarray}\label{a2}
\begin{array}{cc}
{\mathcal L}=\frac{1}{2N}f_{\alpha\beta}(q^\mu)\dot q^\alpha\dot q^\beta-NU(q^\mu)
\end{array}
\end{eqnarray}
where $\alpha,\beta=0,1,..,n-1,$  $f_{\alpha\beta}$ is the metric of $n$-dimensional minisuperspace with
signature $(-,+,...,+)$, $q^\alpha$ denotes the local coordinates of minisuperspace
and $U(q^\mu)$ is the particularization of $-\sqrt{h}R^{(3)}(h_{ij})+V(\text{Matter
fields})$.   The canonical momenta conjugate to the
coordinate $q^\alpha$ is
\begin{eqnarray}\label{a3}
\Pi_\alpha=\frac{\partial{\mathcal L}}{\partial\dot q^\alpha}=\frac{1}{N}f_{\alpha\beta}\dot
q^\alpha.
\end{eqnarray}
Hence, the canonical Hamiltonian will be
\begin{eqnarray}\label{a4}
H_c=\Pi_\alpha\dot q^\alpha-{\mathcal L}=N\left[\frac{1}{2}f^{\alpha\beta}\Pi_\alpha\Pi_\beta+U(q^\mu)\right].
\end{eqnarray}
The gauge freedom in choosing the lapse function, $N$, leads us to the super-Hamiltonian
constraint \begin{eqnarray}\label{a5}
{\mathcal H}=\frac{1}{2}f^{\alpha\beta}\Pi_\alpha\Pi_\beta+U(q^\mu)\approx0.
\end{eqnarray}
The canonical quantization of super-Hamiltonian (\ref{a5}) is accomplished in coordinate representation,
$q^\alpha=q^\alpha$, $\Pi_\alpha=-i\partial_\alpha$, and demanding the annihilation
of time independent wave function by Hermitian super-Hamiltonian
\begin{eqnarray}\label{a6}
{\mathcal H}(q^\alpha,-i\partial_\beta)\Psi(q^\mu)=0.
\end{eqnarray}
The WDW equation (\ref{a6}) is independent of an external time parameter
and consequently there is no preferred evolution parameter.

Let us now review the minisuperspace quantum cosmology in the context of QHJ
formalism developed in \cite{New}.
The complex-valued QHJ equation is readily obtained by substituting the polar form of the complex-valued wave function \cite{Complex}
\begin{eqnarray}\label{a7}
\Psi(q^\mu)=e^{iS(q^\mu)},\,\,\,\,\,\,\,q^\mu\,\,\text{and}\,\,S(q^\mu)\in\mathbb{C},
\end{eqnarray}
into the WDW equation (\ref{a6}). Note that the action
function, $S$, and the coordinates of minisuperspace, $q^\mu$, are analytically
extended to the complex values. Substituting (\ref{a7}) into  the WDW equation
(\ref{a6}) gives us a single QHJ equation
\begin{eqnarray}\label{a8}
\frac{1}{2}f^{\alpha\beta}\nabla_\alpha S\nabla_\beta S+U(q^\mu)+Q(q^\mu)=0.
\end{eqnarray}
In the earlier equation $Q$ denotes the complex quantum potential defined by
\begin{eqnarray}\label{a9}
Q=\frac{1}{2i}\square S=-\frac{1}{2}\left(\frac{\square\Psi}{\Psi}-f^{\alpha\beta}\frac{\nabla_\alpha\Psi\nabla_\beta\Psi}{\Psi^2}\right),
\end{eqnarray}
where $\square=f^{\alpha\beta}\nabla_\alpha\nabla_\beta$ is the D'Alembert
operator. Note that
there is no expansion in powers of $\hbar$ in the derivation and Eq.(\ref{a8}) is exact. The QHJ applied to quantum cosmology states that the quantum trajectories
are complex where these observable independent trajectories are given by
QHJ equation (\ref{a8}). The guidance equation, in analogy to standard Bohmian mechanics is given by \cite{New, Complex, Leacock}
\begin{eqnarray}\label{a10}
\Pi_\alpha=\nabla_\alpha S,\,\,\,\,\,\,\,\,\,\,\Pi_\alpha\in\mathbb{C},
\end{eqnarray}
where the complex quantum moment, $\Pi_\alpha$, is related to the velocity
field by \cite{New}
\begin{eqnarray}\label{a11}
\Pi_\alpha=\frac{1}{N}f_{\alpha\beta}\dot q^\beta,\,\,\,\,\,\,\,\,\,\,\,N\in\mathbb{R}.
\end{eqnarray}
 Furthermore, Eqs.(\ref{a8}) and (\ref{a10}) give us the complex quantum super-Hamiltonian
 constraint
 \begin{eqnarray}\label{a13}
 {\mathcal H}=\frac{1}{2}f^{\alpha\beta}\Pi_\alpha\Pi_\beta+\frac{1}{2i}f^{\alpha\beta}\nabla_\alpha\Pi_\beta+U=0,
 \end{eqnarray}
which is a Riccati-type PDE. To obtain complex quantum trajectories we need to solve the earlier partial differential equation.  Also, the invariance of wave function (\ref{a7})
with respect to adding to the action function by an integer
multiple of $2\pi$ and the definition of momenta in (\ref{a10}) gives
\begin{eqnarray}\label{a14}
\oint_C \Pi_\alpha dq^\alpha=2n\pi,\,\,\,\,n=1,2,3,...,
\end{eqnarray}
where $C$ is a counter clockwise contour in the complex configuration space, enclosing the
real line between the classical turning points. The last equation represents
the compatibility of the QHJ equation (\ref{a13}) and the WDW equation (\ref{a6}).
Even though this approach appears similar to  the familiar WKB scheme,
it is worth pointing out that Eq.(\ref{a14}) reproduces the exact quantized energy eigenvalues \cite{See}.

Let us now apply the above formalism of quantum cosmology, as an example,
to the flat FLRW minisuperspace model with perfect fluid as the matter field.
In principle, due to the quantum nature of the model,
the matter content should be described by some sorts of fundamental fields, as done in \cite{CC}. However, general
exact solutions are hard to find at the presence of fundamental fields and the Hilbert space structure
is obscure and it is a subtle matter to recover the notion of a semiclassical time \cite{CC,Ish}.  It is clear that using perfect fluid is essentially semiclassical from the start. But it has the advantage of introducing a variable, connected with the matter degrees of freedom which can naturally be identified with time, leading to a well-defined Hilbert space structure \cite{Shu}. Another  attractive feature of the
phenomenological description of matter degree of freedom is that it allows us to treat the barotropic equation of state. 
\section{QHJ cosmology with perfect fluid}
Let us consider now the Robertson-Walker spatially  flat metric
\begin{eqnarray} \label{lineelement}
ds^{2}=-N^{2}(t)dt^{2}+a^{2}(t)(dx^{2}+dy^{2}+dz^{2}).
\end{eqnarray}
The action functional corresponding to the model consists of a gravitational part and a matter part (which is considered as a perfect fluid) is given by \cite{Haw}
\begin{eqnarray} \label{action}
\mathcal{A}=\frac{1}{16\pi G}\int{R\sqrt{-g}d^{4}x}-\int{\rho\sqrt{-g}d^{4}x},
\end{eqnarray}
where $\rho=\sum\rho_i$ is the total energy density of fluid with components
$\rho_i$.
To obtain the correct dynamical equations from a
variation of above action functional, it is necessary to require
the current vector of the fluid to be covariantly conserved. For a noninteracting perfect fluid components, with barotropic equation of state $p_{i}=\omega_{i}\rho_{i}$, the covariant conservation of current vector field gives $\rho_{i}=\rho_{0i}(\frac{a}{a_{0}})^{-3(1+\omega_{i})}$, where $\omega_{i}$ denotes the equation of state parameter of the $i$-th component of fluid and  the subscript zero represents the values at the measuring time epoch. To simplify the action functional we define a new dimensionless time coordinate by $\eta=H_{0}t$, where $H_0$ is the Hubble parameter in the comoving frame. Also,
we define a dimensionless scale factor as $x=\frac{a}{a_{0}}$ and rewrite the energy densities in terms of the cosmological density parameters $\Omega_{i}=\frac{8 \pi G}{3H^{2}_{0}}\rho_{0i}$ at the measurement epoch. A straightforward calculation shows that the Lagrangian of the model, up to a multiplicative constant $\frac{3V_3 H_0a_0^3}{4\pi G}$ where $V_3$ is the spacial volume of 3-metric, will be
\begin{eqnarray}\label{lagrangian}
\mathcal{L}=-\frac{1}{2}\Big{(}\frac{\dot{x}^2}{\tilde{N}}+\tilde{N}\sum_{i}{\Omega_{i}x^{1-3\omega_{i}}}\Big{)},
\end{eqnarray}
where we defined the lapse function in the conformal frame as $\tilde{N}=\frac{N}{x}$ and overdot denotes differentiation respect to $\eta$. The conjugate momenta to the dimensionless scale factor, $x$, and the lapse function, $\tilde{N}$, are given by
\begin{eqnarray} \label{momentoms}
\Pi_{x}=\frac{\partial{\mathcal{L}}}{\partial{\dot{x}}}=-\frac{\dot{x}}{\tilde{N}},\,\,\,\,\,\,\,
\Pi_{\tilde{N}}=\frac{\partial{\mathcal{L}}}{\partial{\dot{\tilde{N}}}}=0.
\end{eqnarray}
The canonical Hamiltonian turns out to be
\begin{eqnarray}\label{b19}
H_c=\tilde{N}\left(-\frac{\Pi_{x}^{2}}{2}+\frac{1}{2}\sum_{i}\Omega_{i}x^{1-3\omega_{i}}\right).
\end{eqnarray}
The gauge freedom on choosing the lapse function gives us the super-Hamiltonian
constraint\begin{eqnarray}\label{b20}
\mathcal H=-\frac{\Pi_{x}^{2}}{2}+\frac{1}{2}\sum_{i}\Omega_{i}x^{1-3\omega_{i}}\approx0.
\end{eqnarray}
Also, the classical field equation is
\begin{eqnarray}\label{b21}
\frac{d}{d\eta}\left(\frac{\dot x}{\tilde N}\right)-\frac{\tilde N}{2}\sum_i\left(1-3\omega_i\right)\Omega_ix^{-3\omega_i}=0.
\end{eqnarray}
At the quantum lever, the super-Hamiltonian (\ref{b20}) by promoting
the canonical variables to operators gives the corresponding WDW equation
\begin{eqnarray}\label{n1}
\frac{d^2\Psi(x)}{dx^2}+\sum_{i}\Omega_{i}x^{1-3\omega_{i}}\Psi(x)=0.
\end{eqnarray}
\subsection{QHJ cosmology with one component
perfect fluid}
Let us  first consider a very simple model where the fluid has only one
component.
In this case, the classical solution of field Eqs.(\ref{b20}) and (\ref{b21})
in  gauge $\tilde N=1$ is given by
\begin{eqnarray}\label{b22}
x=\left(\frac{\sqrt{\Omega}}{2}(3\omega+1)\eta\right)^{\frac{2}{3\omega+1}}.
\end{eqnarray}
The WDW equation (\ref{n1}) in this case will be reduce to
\begin{eqnarray}\label{b23}
\frac{d^2}{dx^2}\Psi(x)+\Omega x^{1-3\omega}\Psi(x)=0.
\end{eqnarray}
Note that
the domain of definition of dimensionless scale factor is, $x\in(0,\infty)$, hence the
super-Hamiltonian operator, $\mathcal H$, is defined on a dense domain $D(\mathcal H)=C_0^\infty(0,\infty)$. Then, the super-Hamiltonian is Hermitian (or symmetric) if the wave function
satisfy the DeWitt boundary condition \cite{DeWitt}
\begin{eqnarray}\label{b24}
\Psi(0)=0.
\end{eqnarray}
The solution of WDW equation (\ref{b23}) with boundary condition (\ref{b24})
is the Bessel function of the first kind
\begin{eqnarray}\label{b25}
\Psi(x)={\mathcal N}\sqrt{x}J_\frac{1}{3(1-\omega)}\left(\frac{2\sqrt{\Omega}}{3(1-3\omega)}x^\frac{3(1-\omega)}{2}\right).
\end{eqnarray}
The wave function (\ref{b25}) is real and consequently in the Bohmian cosmology
\cite{Bohmian} the scale factor does not have dynamics.

Let us investigate the QHJ formalism of example model. At the first step we need to extend
the coordinate $x$ to the complex domain, $x=x_\text{R}+ix_\text{I}\in\mathbb{C}$.
Inserting the ansatz (\ref{a7}) into the WDW equation (\ref{b23}) gives
the complex-valued QHJ equation
 \begin{eqnarray}\label{b26}
-\Pi^2_x+i\frac{d\Pi_x}{dx}+\Omega x^{1-3\omega}=0,
\end{eqnarray}
with boundary condition (equivalent to the DeWitt boundary condition
(\ref{b24})) $\Pi_x(0)=\infty$. Now, instead of
the WDW equation (\ref{b23}) we deal with the complex quantum super-Hamiltonian constraint (\ref{b26}). The solution of (\ref{b26}) is
\begin{eqnarray}\label{b27}
\Pi_x=i\left(\frac{J_\frac{4-3\omega}{3(1-\omega)}\left(\frac{2\sqrt{\Omega}}{3(1-\omega)}x^\frac{3(1-\omega)}{2}\right)}{J_\frac{1}{3(1-\omega)}\left(\frac{2\sqrt{\Omega}}{3(1-\omega)}x^\frac{3(1-\omega)}{2}\right)}\sqrt{\Omega}x^\frac{1-3\omega}{2}-\frac{1}{x}  \right),
\end{eqnarray}
where the complex quantum momenta, $\Pi_x$, is related to the velocity
by equation (\ref{momentoms}) or equivalently
\begin{eqnarray}\label{b28}
\Pi_x=-\frac{1}{\tilde N}\frac{dx}{d\eta}.
\end{eqnarray}
 However, an analytical treatment of (\ref{b27}) and (\ref{b28}) does not seem feasible. Therefore,
 we use the following asymptotic forms  of Bessel function of first kind
 for complex argument
\begin{eqnarray}\label{b29}
J_\nu(z)=\begin{cases}
\frac{1}{\Gamma(\nu+1)}\left(\frac{z}{2}\right)^\nu,\hspace{.5cm}0<|z|\ll\sqrt{1+\nu},\\
\frac{1}{\sqrt{2\pi z}}\exp{\left(i(z-\frac{\nu\pi}{2}-\frac{\pi}{4})\right)},\,\,\,|z|\gg1.
\end{cases}
\end{eqnarray}
The approximate solution of Eqs.(\ref{b27}) and (\ref{b28})
in the gauge $\tilde N=1$ then will be
\begin{eqnarray}\label{b30}
\begin{cases}
x_\text{R}=-x_\text{I}=\sqrt{\eta}, \hspace{1.5cm}0<|x|\ll \sqrt{\frac{4-3\omega}{3(1-\omega)}},\\
x_\text{I}=0, \,\,\,\,\,\,x_\text{R}=\left(\frac{(3\omega+1)\sqrt{\Omega}}{2}\eta\right)^{\frac{2}{3\omega+1}},\,\,\,\,\,|x|\gg1.
\end{cases}
\end{eqnarray}
These solutions show that at the very early universe the complex scale factor grows
 with the square root of conformal time (like a  stiff matter or massless
scalar field) when the quantum effects are dominated.
This behavior
is independent to the type of perfect fluid and the
origin of  stiff matter is quantization.  On the other hand,
for  large values of scale factor, $|x|\gg1$, the imaginary part of scale factor vanishes and we have a emerged classical universe.
\subsection{QHJ cosmology of a universe made of dust and radiation}
As a second example let us  consider a fluid of non interacting dust,
$\omega_d=0$, and radiation, $\omega_r=\frac{1}{3}$.
The classical solution of field Eqs.(\ref{b20}) and (\ref{b21}) is given
by \begin{eqnarray}\label{b31}
x=\frac{\Omega_d}{4}\eta^2+\sqrt{\Omega_r}\eta,
\end{eqnarray}
where $\Omega_r$ and $\Omega_r$ are the density parameters of dust and radiation,
respectively.
 The WDW equation (\ref{n1})
for this case is
\begin{eqnarray}\label{b32}
\frac{d^2}{dx^2}\Psi(x)+\left(\Omega_dx+\Omega_r\right)\Psi(x)=0.
\end{eqnarray}
The solution of equation (\ref{b32}) with boundary condition (\ref{b24})
is
\begin{eqnarray}\label{b33}
\Psi(x)={\mathcal N}Ai\left(-(\Omega_dx+\Omega_r)\Omega_d^{-\frac{2}{3}}\right),
\end{eqnarray}
where $Ai(x)$ is the Airy function of first type and ${\mathcal N}$ denotes
the normalization constant. Similar to the first example, the real value
 wave function (\ref{b33}) implies that the Bohmian momenta vanishes and consequently the Bohmian trajectories just sit at one place.

Let us investigate the QHJ of    model. Similar to the previous example we  extend
the coordinate $x$ to the complex domain, $x=x_\text{R}+ix_\text{I}\in\mathbb{C}$. Now, instead of the
WDW equation (\ref{b32}) we deal with the complex quantum super-Hamiltonian constraint (\ref{a13}), which is given by the following equation
\begin{eqnarray}\label{b34}
-\Pi^2_x+i\frac{d\Pi_x}{dx}+\Omega_d x+\Omega_r=0,
\end{eqnarray}
where the boundary condition, equivalent to the DeWitt boundary condition
(\ref{b24}), is $\Pi_x(0)=\infty$.
The solution of this Riccati-type equation is
\begin{eqnarray}\label{b35}
\Pi_x=i\Omega_d^\frac{1}{3}\frac{Ai'(u)}{Ai(u)},
\end{eqnarray}
where $u=-(\Omega_dx+\Omega_r)\Omega_d^{-\frac{2}{3}}$,  $Ai'(u)=\frac{dAi(u)}{du}$
and the complex quantum momenta is related to the velocity
by equation (\ref{b28}).
Using the DeWitt boundary condition in wave function (\ref{b33}), or equivalently
$\Pi(0)=\infty$ at (\ref{b35}) we obtain,
$-\Omega_r\Omega_d^{-\frac{2}{3}}=a_n$,
where $a_n$ are the roots of Airy function. For large values of $n$
the roots are given by $a_n=-\left(\frac{3\pi}{2}(n-\frac{1}{4})\right)^\frac{2}{3}$
where $n$ is a integer\,.
This gives us the quantization condition for the ratio of density parameters
\begin{eqnarray}\label{b36}
\Omega_r\Omega_d^{-\frac{2}{3}}=\left(\frac{3\pi}{2}(n-\frac{1}{4})\right)^\frac{2}{3}.
\end{eqnarray}
An analytical treatment of equations (\ref{b28}) and (\ref{b35})  does not seem feasible and
 we use the following  asymptotic form of the Airy function of first type
\begin{eqnarray}\label{b37}
Ai(u)=
\begin{cases}
\frac{dAi(\xi)}{d\xi}|_{\xi=a_n}\Omega_d^\frac{1}{3}x,\,\,\,\,\,\,\,\,|x|\ll1,\\
\\
u^{-\frac{1}{4}}\exp{(-\frac{2i}{3}u^\frac{3}{2})},\,\,\,\,|x|\gg1,
\end{cases}
\end{eqnarray}
where $u=-(\Omega_dx+\Omega_r)\Omega_d^{-\frac{2}{3}}$.
Thus, the approximate solution of equations (\ref{b28}) and (\ref{b35})
in the gauge $\tilde N=1$ will be
\begin{eqnarray}\label{b38}
\begin{cases}
x_\text{R}=x_\text{I}=\sqrt{\eta},\hspace{2.2cm}|x|\ll1,\\
x_\text{I}=0,\,\,x_\text{R}=\frac{\Omega_d}{4}\eta^2+\sqrt{\Omega_r}\eta,\,\,\,|x|\gg1.
\end{cases}
\end{eqnarray}
The above solution shows that in the very early universe, where the quantum effects are dominated, the scale factor is complex and it grows as  the scale
factor of a
universe filled with massless scalar field. On the other hand,
for  large values of scale factor, the imaginary part of scale factor vanishes and the motion
is entirely classical. The solution for $|x|\gg1$ at measuring time $\eta_0$
gives $\Omega_d+\Omega_r=1$. Therefore, using this condition and equation
(\ref{b36}) we can obtain the density parameters of radiation and dust in
terms of quantum number $n$. For very large values of $n$ we have
\begin{eqnarray}\label{b39}
\begin{cases}
\Omega_d\simeq\frac{2}{3\pi n},\hspace{1cm}\Omega_r\simeq1,\\
x_\text{R}=\eta,\hspace{1.4cm}x_\text{I}=0
\end{cases}
\end{eqnarray}
Hence, the emerged universe is completely classical with real
minisuperspace, radiation dominated and the density of dust is lower by many orders of magnitude.
\subsection{QHJ cosmology of a universe made of stiff matter and
radiation}
Finally, let us  consider a fluid of
non interacting stiff matter, $\omega_s=1$, and radiation, $\omega_r=\frac{1}{3}$.
The classical solution of field equations (\ref{b20}) and (\ref{b21}) is
\begin{eqnarray}\label{s1}
x^2=\Omega_r\eta^2+2\sqrt{\Omega_s}\eta,
\end{eqnarray}
where $\Omega_s$ is the density parameter of stiff matter. The WDW equation (\ref{n1}) in this case will be reduce to
\begin{eqnarray}\label{s2}
\frac{d^2\Psi(x)}{dx^2}+\left(\Omega_r+\Omega_sx^{-2}\right)\Psi(x)=0,
\end{eqnarray}
where the solution assigned by the DeWitt boundary condition (\ref{b24})  is
\begin{eqnarray}\label{s3}
\Psi(x)=\mathcal N\sqrt{x}J_{\frac{1}{2}\sqrt{1-4\Omega_s}}\left(\sqrt{\Omega_r}x\right).
\end{eqnarray}
To study the QHJ cosmology of model, we extend the coordinate $x$
to the complex domain, $x = x_\text{R} + ix_\text{I}\in\mathbb{C}$. The complex quantum
super-Hamiltonian constraint (\ref{a13})  will be
\begin{eqnarray}\label{s4}
-\Pi^2_x+i\frac{d\Pi_x}{dx}+\Omega_s x^{-2}+\Omega_r=0.
\end{eqnarray}
The solution of (\ref{s4}) is
\begin{eqnarray}\label{s5}
\Pi_{x}=i\left(\frac{\sqrt{\Omega_r}J_{1+\nu}\left(\sqrt{\Omega_r}x\right)}{J_{\nu}\left(\sqrt{\Omega_r}x\right)}-\frac{\nu+1}{x}\right),
\end{eqnarray}
where $\nu=\frac{\sqrt{1-4\Omega_s}}{2}$.
Using asymptotic forms of the Bessel function defined in (\ref{b29}), Eq.(\ref{s5})
will reduce to
\begin{eqnarray}\label{s6}
\Pi_x=\begin{cases}
-\frac{i(2\nu+1)}{2x},\hspace{.5cm}0<|x|\ll\sqrt{1+\nu},\\
-\sqrt{\Omega_r},\hspace{1.5cm}|x|\gg1.
\end{cases}
\end{eqnarray}
Therefore, the asymptotic solution of  (\ref{b28}) in the gauge $\tilde N=1$ is given by
\begin{eqnarray}\label{s7}
\begin{cases}
x_R=x_I=\sqrt{\frac{1+2\nu}{2}}\sqrt{\eta},\hspace{1.1cm}|x|\ll\sqrt{1+\nu},\\
x_I=0,\,\,x_R=\sqrt{\Omega_r}\eta,\hspace{1.3cm}|x|\gg1.
\end{cases}
\end{eqnarray}
Asymptotic form of the scale factor in (\ref{s7}) shows that, in the very early universe the scale factor is complex and stiff matter
dominated, similar to the previous couple of model examples. On the other hand, for late times the classical radiation dominated
universe emerges with a real scale factor.

 Let us now examine the effect
of factor ordering on the behaviour of model.
The WDW equation (\ref{n1}) in general factor ordering  given by
\cite{Po}
\begin{eqnarray}\label{ordering}
\Pi^2_x=-\frac{1}{2}\left(x^\alpha\partial_xx^\beta\partial_xx^\gamma+x^\gamma\partial_xx^\beta\partial_xx^\alpha\right),
\end{eqnarray}
with, $\alpha+\beta+\gamma=0$, will be
\begin{eqnarray}\label{deformed}
\frac{d^2\Psi(x)}{dx^2}+\left(qx^{-2}+\sum_i\Omega_ix^{1-3\omega_i}\right)\Psi(x)=0,
\end{eqnarray}
where $q=(\alpha+\gamma+2\alpha\gamma)/2$ denotes the ordering parameter.
Also, the corresponding QHJ equation (\ref{a13}) will be
\begin{eqnarray}\label{bb26}
-\Pi^2_x+i\frac{d\Pi_x}{dx}+qx^{-2}+\sum_i\Omega_i x^{1-3\omega_i}=0.
\end{eqnarray}
 Hence,  factor ordering only changes the density parameter
of stiff matter defined in the WDW equation (\ref{s2}) or equivalently in complex quantum
super-Hamiltonian constraint (\ref{s4}) as $\bar\Omega_s=\Omega_s+q$. But,
in previous examples it has been shown that the stiff matter content of universe
is not contributed at the semi-classical region. Therefore, the factor ordering
is not contributed at the behaviour of emerged classical universe. Also, in canonical quantum cosmology,
it has also been stated that the factor ordering  is not very important to the
theory as a whole \cite{1113}, namely, from a semiclassical
perspective \cite{1415}.

Let us  investigate the behavior of the model for very early universe.
As illustrated
in all above
examples  the dimensionless scale factor for very early universe  is proportional with the roots square of dimensionless conformal time, $x_\text{R}\simeq\sqrt\eta$. Using this relation on can easily show that the scale
factor and Hubble parameter for all of the example early universes (when
the quantum gravity effects are dominated) in comoving
frame are given by
\begin{eqnarray}\label{com}
a(t)\simeq  a_0\left({H_0t}\right)^\frac{1}{3},\,\,\,\,\,\,
H=\frac{1}{3t},
\end{eqnarray}
 where $t$ is the comoving time related to the $\eta$ by $dt=\frac{x_\text{R}}{H_0}d\eta$,
  $a_0$ and $H_0$ denote the scale factor and Hubble parameter at the classical
measuring epoch. Hence the value of scale factor and Hubble parameter at the Planck time
 are given by
 \begin{eqnarray}\label{PPl}
a_\text{P}\simeq a_0(H_0t_\text{P})^\frac{1}{3},\,\,\,\,\,\,\, H_\text{P}\simeq\frac{1}{t_\text{P}},
 \end{eqnarray}
where $t_\text{P}$ denotes the Planck time. Setting the measuring  time as the GUT time, $t_0=t_\text{GUT}$, also using the  values of the Hubble parameter
and the scale factor at GUT time, $H_\text{GUT}t_\text{P}\simeq10^{-8}$, $a_0=a_\text{GUT}\simeq
2$ cm, \cite{Roos} we obtain the linear size
of universe at the Planck time
\begin{eqnarray}\label{p12}
a_\text{P}\simeq a_\text{GUT}10^{-2}\simeq 1 \text{mm}.
\end{eqnarray}
In other words, for a scale factor smaller than $a_\text{p}$ (belove the Planck
time) the quantum effects
are dominated and the scale factor
is complex while for $a\geq a_\text{P}$ the emerged
universe is completely  classical with real minisuperspace. 

{In canonical quantum cosmology, it is generally believed that it is impossible to obtain a large classical universe from quantum cosmology
without an inflationary phase in the classical expanding era, because the
assumption  is that the typical linear  size of the universe after
leaving the quantum regime should be around the Planck length, and therefore, the  decelerated classical
expansion after quantum regime is not sufficient to enlarge the universe in the time available. More precisely, suppose that
the spacetime metric is the spatially closed, homogeneous and isometric and
the matter fields contribution to the WDW equation separate from the gravitational
contribution, like the conformal scalar
field \cite{Conformal} or the Yang-Mills radiation field \cite{YM}. }
Then the wave function of universe can be written
\begin{eqnarray}\label{wave}
\Psi_n=\psi_n(a)\phi_n(\text{matter}),
\end{eqnarray}
where $\phi_n$ represents the matter part of the wave function and $\psi_n$ represents the gravitational part which is the $n$-th
harmonic oscillator wave function. If we assume that linear size of the universe after leaving the quantum regime is around the Planck length, 
both the matter field and the gravitational degree of freedom have small quantum numbers since the two degrees of freedom are connected.
 Therefore,  when the linear size of universe is confined to
about a Planck length both the matter field and the gravitational degree of freedom have small quantum numbers. On the other hand, in order to get
a Standard FRLW universe, enormously large excitation quantum numbers are required  ($n\simeq10^{120}$), which makes
a quantum description irrelevant and at the same time shows once more the
unnaturalness of  the conditions for the present universe as seen from the quantum point of view  \cite{YM}. 
On the contrary,  as we showed in \cite{New}, in quantum Hamilton-Jacobi
approach applied to quantum cosmology with conformally coupled scalar field, the typical linear size of universe after leaving the quantum regime
is around a millimeter which is clearly consistence with large
values of the quantum numbers mentioned above. In this paper also we considered that the universe is spatially
flat and consequently there is not exist any discrete quantum number. However, equation (\ref{p12})  shows that at the end of quantum
era the linear size of universe is very large and consequently it is possible to have a sufficiently big universe emerged from the quantum dominated universe,
without needing any classical inflationary phase afterwards
to make it grow to its present size.

\section{Conclusions}
{In this paper the quantum Hamilton-Jacobi cosmology of spatially flat homogeneous and isotropic early universe has been studied whose matter content is the perfect fluid.}
 The idea of exact classical-quantum correlation is introduced in the sense that the WDW equation reduces
to the quantum Hamilton-Jacobi equation in cosmology under the
substitution $\Psi=\exp(iS)$. The resulting model is free of problems of
Copenhagen
interpretation of quantum cosmology such as collapsing of wave function,
probabilistic  interpretation of wave function and
time problem.  We show that at the very early universe (back beyond one Planck time) we have an evolving complex universe. On
the other hand,  the emerged universes around one Planck time are coincide with the corresponding
original classical universes, without needing any classical inflationary phase afterwards
to make it grow to its present size.
 This paper follows the same procedure which was done in the previous paper \cite{New}, by the authors. We
are aware that our results are obtained within a very simple cosmological model. A wider analysis, with
less restrictive cosmologies and other matter fields, should follow. Using instead, e.g., a scalar
and Yang-Mills fields would be more generic and more realistic from the
point of view of matter interaction with the gravitational
field in a high energy regime, where
quantum effects can be expected. We believe the above results offer an
insight into the relation between classical and quantum cosmological models and that the particular simple models that we have
studied may serve as useful starting point for more ambitious investigation.



\begin{thebibliography}{}
%
%
\bibitem{New} Fathi, M., Jalalzadeh, S.,  Moniz, P.V.: Eur. Phys. J. C {\bf76},
 527 (2016), [arXiv:1609.04488]
\bibitem{Bohmian}   Alvarenga, F.G.,  Fabris, J.C.,  Lemos, N.A., Monerat, G.A.: Gen. Rel. Grav. {\bf34}, 651 (2002), [arXiv:gr-qc/0106051];
Shojai, F., Shirinifard, A.: Int. J. Mod. Phys. D {\bf14}, 1333 (2005),
[arXiv:gr-qc/0504138];  Pinto-Neto, N.,  Fabris, J.C.: Class. Quantum Grav. {\bf30}, 143001 (2013),
[arXiv:1306.0820];  Pedram, P., Jalalzadeh, S: Phys. Lett. B {\bf660}, 1 (2008), [arXiv:0712.2593];  Peter, P., Pinto-Neto, N.: Phys. Rev. D {\bf78}, 063506 (2008), [arXiv:0809.2022]
\bibitem{Complex}  Chou, C.C.,   Wyatt, R.E.: Phys. Rev. A {\bf76}, 012115
(2007); Gozzi, E.: Phys. Lett. B {\bf165}, 351 (1985);  Bhalla, R.S.,  Kapoor,
A.K.,  Panigrahi, P.K.: Am. J. Phys. {\bf65}, 1187 (1997)
\bibitem{Leacock}  Leacock, R.A.,  Padgett, M.J.: Phys. Rev. Lett. {\bf50}, 3 (1983);  Phys. Rev. D {\bf28}, 2491 (1983);
 Am. J. Phys. {\bf55}, 261 (1986)
 \bibitem{Gozzi} Gozzi, E.: Phys. Lett. B {\bf165}, 351 (1985)
 \bibitem{John} John, M.V.: Found. Phys. Lett. {\bf 15}, 329 (2002), [arXiv:quant-ph/0109093]; Gravitation and Cosmology, {\bf21}, 208  (2015), [arXiv:1405.7957]
\bibitem{R}  Bhalla, R.S.,  Kapoor, A.K., Panigrahi, P.K.: Am. J. Phys. {\bf65},
1187 (1997);   Bhalla, R.S.,  Kapoor, A.K., and  Panigrahi, P.K.: Mod. Phys. Lett. A {\bf12}, 295 (1997)
\bibitem{1724} Yang, C-D.: Ann. Phys. (N.Y.) {\bf319}, 399 (2005);
Ann. Phys. (N.Y.), {\bf319}, 444 (2005); Int. J. Quantum Chem. {\bf106},
1620 (2006);  Ann. Phys. (N.Y.) {\bf321}, {2876} (2006)
 \bibitem{Har}   Yang, C-D.: Phys.  Lett. A {\bf372}, 6253 (2008); Chaos, Soliton Fract. {\bf30}, 342 (2006)
\bibitem{See}  Sree Ranjani, S.,  Geojo, K.G.,  Kapoor, A.K., Panigrahi,
P.K.: Mod. Phys. Lett. A  {\bf19}, 1457 (2004)
\bibitem{CC} Kiefer, C.: Phys. Rev. D {\bf38}, 1761 (1988)
\bibitem{Ish}  Isham, C.J.: [arXiv:gr-qc/9210011]
\bibitem{Shu} Gotay, M.J., Demaret, J.: Phys. Rev. D {\bf28}, 2402 (1983);  Alvarenga, F.G.,  Fabris, J.C.,  Lemos, N.A., Monerat, G.A.: Gen.
Rel. Grav. {\bf34}, 651 (2002), [arXiv:gr-qc/0106051];  Pedram, P., Mirzaei,
M., Jalalzadeh, S.,  Gousheh, S.S.: Gen. Rel. Grav. {\bf40}, 1663
(2008), [arXiv:0711.3833]; Vakili, B.: Class. Quantum. Grav. {\bf27}, 025008
(2010), [arXiv:0908.0998]; Pedram, P., Jalalzadeh, S.,  Gousheh,  S.S.:
Phys. Lett. B. {\bf655}, 91 (2007), [arXiv:0708.4143]; Pedram, P., Jalalzadeh
S.,  Gousheh, S.S.: Class. Quantum. Grav. {\bf24}, 5515 (2007), [arXiv:0709.1620]
\bibitem{Haw}  Hawking, S.W.,  Ellis, G.F.R., {\it The Large Scale Structure
of Space-Time} (Cambridge University Press, Cambridge,
United Kingdom, 1973)
\bibitem{DeWitt} DeWitt, B.S.: Phys. Rev. {\bf160}, 1113 (1967)
\bibitem{Po}  Pedram, P.,  Jalalzadeh, S.: Phys. Rev. D {\bf 77}, 123529 (2008), [arXiv:0805.4099]
\bibitem{1113} Vilenkin, A.: Phys. Rev. D {\bf58}, 067301 (1998)
\bibitem{1415} Bojowald, M., Simpson, D.: Class. Quantum Grav. {\bf31}, 185016 (2014); Kiefer, C.: Lect. Notes Phys. {\bf434}, 170
(1994); Louko, L.: Ann. Phys. (N.Y.) {\bf181}, 318 (1988)
\bibitem{Roos} Roos M.: {\it Introduction to Cosmology}. Wiley, West Sussex,
 England (2003)
\bibitem{Conformal} Bertolami, O.,  Mour\~ao, J.M.: Class. Quantum Grav. {\bf8}, 1271 (1991); Jalalzadeh, S.,  Rostami, T., Moniz, P.V.: Int. J. Mod. Phys. D {\bf25} 1630009 (2016)
\bibitem{YM} Cavagli\`a, M., de Alfaro, V.: Mod. Phys. Lett. A {\bf9}, 569 (1994)
\end{thebibliography}


\end{document}